\documentclass[aps,prb,showpacs,twocolumn,amsmath,amssymb,superscriptaddress]{revtex4}
\usepackage{graphicx}
\usepackage{Jonasmacros}
\usepackage{bm}
\begin{document}
\date{\today}
\title{Vibrational coherence in electron spin resonance in nanoscale oscillators}
\author{J. Fransson}
\email{Jonas.Fransson@fysik.uu.se}
\affiliation{Department of Physics and Materials Science, Uppsala University, Box 530, SE-751 21\ \ Uppsala}
\author{Jian-Xin Zhu}
\email{jxzhu@lanl.gov}
\affiliation{Theoretical Division, Los Alamos National Laboratory, Los Alamos, New Mexico 87545, USA}

\begin{abstract}
We study a scheme for electrical detection, using electron spin resonance, of coherent vibrations in a molecular single electron level trapped near a conduction channel. Both equilibrium spin-currents and non-equilibrium spin- and charge currents are investigated. Inelastic side-band anti-resonances corresponding to the vibrational modes appear in the electron spin resonance spectrum. 
\end{abstract}
\maketitle

Inelastic effects arising due to coupling between charge carriers and local vibrational modes (vibrons) in nanoscale electronics devices has gained an enormous interest recently. Peaks and dips often observed in the differential conductance of molecular electronics devices\cite{andres1996} may indicate strong effects from electron-vibron coupling. Effects from vibrons have been investigated\cite{wingreen1989,zhu2003} in molecular quantum dots and single electron transistor, in Josephson junctions,\cite{gorelik2001} and on surfaces using scanning tunneling microscopy (STM).\cite{stipe1998}

The interplay between vibrons and charge carrier is expected to generate dynamical signatures also in the spin current or spin-dependent transport. Such dynamics should consequently be observable in electron-spin resonance (ESR),  which will thus allow for electrical detection of both spin and vibron modes. In this paper, we apply the ESR set-up~\cite{durkan2004,martin2003} to a molecular quantum dot with electron levels coupled to vibrons, and we show the emergence of anti-resonances in the spin-current at frequencies equal to integral numbers of the vibrational mode. The anti-resonances can be explained as interference between opposite spin tunneling electron wave functions traversing different molecular excitations. Such information would be useful not only to the conventional semi-conductor industry, but also to novel research directions such as spintronics and molecular electronics. Distinct from earlier work on ESR set-up,\cite{martin2003} we are addressing the novel signatures  arising from the electron-vibron coupling, which will manifest in both equilibrium spin current and nonequilibrium spin dependent transport.

We model the resonator to be oscillating with frequency $\omega_0$,  where the vibrational motion is weakly coupled to the electrons with strength $\lambda$. We consider the dynamics of a single molecular level $\dote{0}$ coupled to external thermal baths. The level is spin split by the external magnetic field $B_0$, $\dote{\down}-\dote{\up}=\omega_r\equiv g\mu_BB_0$, where $g$ and $\mu_B$ are the gyromagnetic ratio and Bohr magneton, respectively. The spins are coupled by a rotating magnetic field $B_1(\cos{\omega_1t},\sin{\omega_1t})$ applied perpendicular to $B_0$, and we assume $2g\mu_BB_1\ll\omega_0, \omega_r$. We employ the model $\Hamil=\Hamil_c+\Hamil_d+\Hamil_T$, where
\begin{eqnarray}
\Hamil_d&=&
	\sum_\sigma[\dote{\sigma}+\lambda(a^\dagger+a)+Un_{\bar\sigma}/2]n_\sigma
\nonumber\\&&
	-g\mu_BB_1(\ddagger{\up}\dc{\down}e^{i\omega_1t}
						+\ddagger{\down}\dc{\up}e^{-i\omega_1t})
	+\omega_0a^\dagger a,
\end{eqnarray}
describes the molecular states, while $\Hamil_c=\sum_{k\sigma}\dote{k}n_{k\sigma}$ and $\Hamil_T=\sum_{k\sigma}v_k\cdagger{k}\dc{\sigma}+H.c.$, are the Hamiltonians for the bath and the tunneling, respectively. Here $\cdagger{k}$ and $\ddagger{\sigma}$ create an electron with spin $\sigma=\up,\down$ in the bath and molecule, respectively,  $n_{k\sigma}=\cdagger{k}\cc{k}$ and $n_\sigma=\ddagger{\sigma}\dc{\sigma}$. The operators $a^\dagger$ and $a$ denote creation and destruction of the vibrational mode.

We transform the system into the rotating reference frame of the magnetic field through
$\Hamil_{rf}=e^{S_{rf}}\Hamil e^{-S_{rf}}+i(\dt e^{S_{rf}})e^{-S_{rf}}$,
with the unitary transformation $S_{rf}=-i(\omega_1 t/2)[n_\down-n_\up+\sum_k(n_{k\down}-n_{k\up})]$, 
in order to eliminate the time-dependence from the Hamiltonian at the cost of introducing a shift in the electronic energies, i.e. $\leade{k}^{rf}=\dote{k}+\sigma\omega_1/2$ and $\dote{\sigma}^{rf}=\dote{\sigma}+\sigma\omega_1/2$, where the factor $\sigma=\pm1$. The spin-split of the conduction channel electron energies originates from the magnetic pumping field through the hybridization between the localized level and the conduction band. The pumping propagates energy from the molecule to the conduction channel and generate the spin chemical potentials $\mu_\sigma=-\sigma\omega_1/2$ (with reference to  $\dote{F}=0$) in the conduction channel. The frequency of the oscillating magnetic field can thus be regarded as the (spin) bias applied to the system. Despite the spin-imbalance, however, the charge chemical potential is still $\mu=(\mu_\up+\mu_\down)/2=0$.\cite{veillette2004}

Although the system itself is to be considered in equilibrium, the one photon imbalance between the spin channels generates a non-equilibrium condition for the two spin projections of the electrons. An electron in the spin down channel can thus tunnel into the local spin down level $\dote{\down}^{rf}$. The rotating magnetic field flips the spin projection of the localized electron and thereby the electron can tunnel into the spin up channel, and a stationary current builds up by repeated tunneling.

The coupling between the vibrational and electronic degrees of freedom is de-coupled by the canonical transformation $\tilde\Hamil=e^{S_{ph}}\Hamil_{rf}e^{-S_{ph}}$ with $S_{ph}=(\lambda/\omega_0)(a^\dagger-a)\sum_\sigma n_\sigma$. Through this transformation the energy levels of the localized states are turned into $\tilde{\epsilon}_\sigma=\dote{\sigma}^{rf}-\lambda^2/\omega_0$, while the charging energy $\tilde{U}=U-2\lambda^2/\omega_0$, and the tunnelling Hamiltonian is changed into $\sum_{k\sigma}(v_{k\sigma}\cdagger{k}\dc{\sigma}X+H.c.)$, where $X=\exp{[-(\lambda/\omega_0)(a^\dagger-a)]}$. In the present study we assume a weak coupling between the electronic and vibrational degrees of freedom, and that the spin-currents through the system are small. It is then justified to neglect narrowing effects on the tunneling between the conduction channel and the molecular level.\cite{hewson1980}

In the atomic limit and $\tilde{U}=0$, the molecule is reduced to a simple driven two-level system. It is characterized by a coherent weight transfer, Rabi oscillations, between the two spin states, which is complete at resonant rotating frequency $\omega_1=\omega_r$. The spin oscillation period of $T=2\pi/\Omega$, where $\Omega=\sqrt{\Delta^2+4(g\mu_BB_1)^2}$ is the Rabi frequency and $\Delta=\omega_1-\omega_r$ denotes the detuning from the resonance. We transform the molecular electron operators by
\begin{equation}
\left(\begin{array}{c} \dc{\up} \\ \dc{\down} \end{array}\right)
=
{\bf u}
\left(\begin{array}{c} \cs{\up} \\ \cs{\down} \end{array}\right),
\quad {\bf u}=\left(\begin{array}{cc} \cos\phi & -\sin\phi \\ \sin\phi & \cos\phi \end{array}\right)\;,
\end{equation}
where $\tan\phi=2g\mu_BB_1/(\Omega-\Delta)$. The molecular electronic states are diagonal in the new representation giving the molecular Hamiltonian $\sum_\sigma E_\sigma\csdagger{\sigma}\cs{\sigma}+\tilde{U}\csdagger{\up}\cs{\up}\csdagger{\down}\cs{\down}$, with $E_\sigma=(\tilde\epsilon_\up+\tilde\epsilon_\down-\sigma\Omega)/2=\dote{0}-\lambda^2/\omega_0-\sigma\Omega/2$.

The spin-$\sigma$ current $I_\sigma$ is preferably written as
\begin{equation}
I_\sigma=\frac{ie}{h}\tr\int\bfGamma_\sigma
	\{f_\sigma(\omega)\bfG^>(\omega)
	+[1-f_\sigma(\omega)]\bfG^<(\omega)\}d\omega,
\label{eq-sc}
\end{equation}
where $\bfGamma_\sigma={\bf u}_\sigma\Gamma$, $\Gamma=2\pi\sum_k|v_k|^2\delta(\omega-\dote{k})$ and 
\begin{equation}
{\bf u}_\up=\tau^y{\bf u}_\down\tau^y,\ 
{\bf u}_\down=\left(\begin{array}{cc}
	\sin^2\phi & \sin\phi\cos\phi\\ \sin\phi\cos\phi & \cos^2\phi
	\end{array}\right),
\end{equation}
where $\tau^y$ is the $y$-component of the Pauli matrix, whereas $f_\sigma(\omega)=f(\omega-\mu_\sigma)$ is the Fermi function for the spin $\sigma$ channel.

The current contains the lesser (greater) Green functions (GFs) $\bfG^{<(>)}=\{G_{\sigma\sigma'}^{<(>)}\}_{\sigma\sigma'}$. They can be calculated using $\bfG^{<(>)}=\bfG^r\bfSigma^{<(>)}\bfG^a$, where e.g. the retarded GF is defined through $G_{\sigma\sigma'}^r(t)=(-i)\theta(t)\av{\anticom{\cs{\sigma}(t)}{\csdagger{\sigma'}(0)}}$ and similarly for the advanced one. The canonical de-coupling procedure of the electron-vibron coupling casts the GF into the product of an electronic and vibronic part as 
\begin{eqnarray}
G_{\sigma\sigma'}^r(t)&=&
	(-i)\theta(t)\av{\anticom{\tilde{c}_\sigma(t)}{\tilde{c}^\dagger_{\sigma'}(0)}}_{el}
		\av{X(t)X^\dagger(0)}_{vib}
\nonumber\\&=&
	\tilde{G}_{\sigma\sigma'}^{r}(t)\av{X(t)X^\dagger(0)}_{vib}.
\end{eqnarray}
with $\tilde{c}_\sigma(t)=e^{i\tilde{\Hamil}_{el}t}\cs{\sigma}e^{-i\tilde{\Hamil}_{el}t}$, $X(t)=e^{i\tilde{\Hamil}_{vib}t}Xe^{-i\tilde{\Hamil}_{vib}t}$. The renormalization factor caused by the electron-vibron coupling is calculated as $\av{X(t)X^\dagger(0)}_{vib}=e^{-\Phi(t)}$, where $\Phi(t)=(\lambda/\omega_0)^2[n_B(1-e^{i\omega_0t})+(n_B+1)(1-e^{-i\omega_0t})]$, with $n_B=(e^{\beta\omega_0}-1)^{-1}$.\cite{mahan1990} We then calculate the electronic GF $\tilde{G}_{\sigma\sigma'}^r$ in the mean field approximation, in which the Kondo resonance effect is neglected. For arbitrary on-site charging energy, its Fourier transform is given by
$\tilde{G}_{\sigma\bar\sigma}^r(\omega)=0$ and $\tilde{G}_{\sigma\sigma}^r=\tilde{G}_\sigma^r$,
 where
\begin{equation}
\tilde{G}_\sigma^r(\omega)=\frac{\omega-E_\sigma-(1-\av{n_{\bar\sigma}})\tilde{U}}
	{(\omega-E_\sigma+i\Gamma/2)(\omega-E_\sigma-\tilde{U})
		+i\av{n_{\bar\sigma}}\tilde{U}\Gamma/2}\;.
\label{EQ:Green_r}
\end{equation}
and $\av{n_\sigma}=\im\int\tilde{G}^<_\sigma(\omega)d\omega/(2\pi)$. We then find 
\begin{equation}
G^r_\sigma(\omega)=e^{-(\lambda/\omega_0)^2(2n_B+1)}\sum_nI_n(z)e^{n\beta\omega_0/2}
	\tilde{G}^r_\sigma(\omega-n\omega_0)\;,
\end{equation}
where $I_n(z)$ is the $n$th modified Bessel function and $z=2[\lambda/\omega_0]^2\sqrt{n_B[n_B+1]}$.

In the case of weak electron-vibron coupling the contributions to the self-energy $\bfSigma$ from the electron-vibron interaction is negligible, hence, the lesser (greater) self-energy can be approximated by
\begin{equation}
\bfSigma^<=if_\up\bfGamma_\up+if_\down\bfGamma_\down\;,\;\; 
	\bfSigma^>=-i\bfGamma_\up-i\bfGamma_\down+\bfSigma^<\;.
\end{equation}
This leads to  the spin-$\sigma$ current  $I_\sigma=e\Gamma^2\int {\cal T}[f_\sigma(\omega)-f_{\bar\sigma}(\omega)]d\omega/h$, where the transmission coefficient
\begin{equation}
{\cal T}=|G_\up^r(\omega)-G^r_\down(\omega)|^2\sin^2\phi\cos^2\phi\;.
\label{eq-T}
\end{equation}
We notice that the transmission ${\cal T}$ is equal for the two spin channels, which is expected in the stationary regime, thus giving the total spin-current $I_s=\sum_\sigma\tau^z_{\sigma\sigma}I_\sigma=2I_\up$. The form of the transmission coefficient suggests that the spin-current can be interpreted as an interference between tunneling electron wavefunctions in the conduction channel, which are coupled by the molecular level.

The interference interpretation is especially appealing in the context of electron-vibron coupling. For simplicity consider the case of vanishing effective charging energy, $\tilde{U}=0\ (U=2\lambda^2/\omega_0)$, although the arguments remain true for arbitrary $\tilde{U}$. Then, the first factor in Eq. (\ref{eq-T}) can be written as (setting $\omega_\sigma^{r/a}=E_\sigma\mp i\Gamma/2$)
\begin{equation}
|G_\up^r-G_\down^r|^2\sim\biggl|\sum_n
	\frac{\Omega I_n(z)e^{n\beta\omega_0/2}}
		{(\omega-\omega^r_\up-n\omega_0)(\omega-\omega^r_\down-n\omega_0)}
	\biggr|^2.
\label{eq-Gud}
\end{equation}
The main ESR peak is given at $\omega_1=\omega_r$ such that $\Omega=2g\mu_BB_1$ and $E_\sigma=\dote{0}-\lambda^2/\omega_0-\sigma g\mu_BB_1$, corresponding to the $n=0$ term in the transmission coefficient. Because of the electron-vibron coupling, additional features in the spin-current are expected to occur at frequencies $\omega_1=\omega_r+n\omega_0$, corresponding to the vibrational side-bands. Due to the destructive interference between tunneling electron waves passing through different conduction (spin) channels, these satellites to the main ESR peak appear as dips in the spin-current rather than peaks.

\begin{figure}[b]
\begin{center}
\includegraphics[width=8cm]{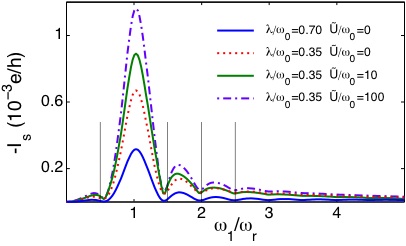}
\end{center}
\caption{(Color online) Equilibrium spin-current dependence of $\lambda$ and $\tilde{U}$. Here, $\omega_r=2$, $g\mu_BB_1=0.2315$, $\Gamma=4\sqrt{5}/25$, and $k_BT=10$, in units of $\omega_0$.}
\label{fig-f1}
\end{figure}

In order to illustrate this argument, we consider the first two terms in Eq. (\ref{eq-Gud}), that is the terms with $n=0,1$
\begin{eqnarray}
\label{eq-T01}
\lefteqn{
\Omega^2[
	I_0^2(z)|\tilde{G}_\up^r(\omega)\tilde{G}_\down^r(\omega)|^2
	+I_1^2(z)|\tilde{G}_\up^r(\omega-\omega_0)
}
\nonumber\\&&\times
	\tilde{G}_\down^r(\omega-\omega_0)|^2
		e^{\beta\omega_0}
	+2\re I_0(z)I_1(z)
		\tilde{G}_\up^r(\omega)\tilde{G}_\down^r(\omega)
\nonumber\\&&\times
			\tilde{G}_\up^a(\omega-\omega_0)\tilde{G}_\down^a(\omega-\omega_0)
		e^{\beta\omega_0/2}
		]
\end{eqnarray}
where the first two terms add positively to the transmission, and peak at $\omega=E_\sigma$ and $\omega=E_\sigma+\omega_0$, respectively. The last term, proportional to
\begin{eqnarray}
	-\re\frac{\Omega/(\omega_0+i\Gamma)}{(\omega-E_\up-\omega_0-i\Gamma/2)(\omega-E_\down+i\Gamma/2)}
\nonumber\\
	+\re\frac{\Omega/(\omega_0+i\Gamma)}{(\omega-E_\down-\omega_0-i\Gamma/2)(\omega-E_\up+i\Gamma/2)}
\label{eq-w01}
\end{eqnarray}
is negligible at $\omega_1=\omega_r$ since then $E_\up\approx E_\down$, which leads to that the two contributions cancel each other. As $\omega_1\rightarrow\omega_r+\omega_0$, on the other hand, we have $E_\up+\omega_0\approx E_\down$, since $\Omega\approx\omega_0$.
Therefore, the first contribution in Eq. (\ref{eq-w01}) roughly equals
\begin{equation}
-\frac{1/[1+(\Gamma/\omega_0)^2]}{(\omega-E_\down)^2+(\Gamma/2)^2}
\label{eq-di}
\end{equation}
while the second contribution is negligible. The expression in Eq. (\ref{eq-di}) peaks around $\omega=E_\down\approx E_\up+\omega_0$, and contributes destructively to the total transmission coefficient in Eq. (\ref{eq-T01}). An estimate of the ratios between the third and first, and third and second terms in Eq. (\ref{eq-T01}) at $\omega_1\approx\omega_r+\omega_0$, yields the lower bounds
\begin{eqnarray}
\biggl|\frac{I_1(z)}{2I_0(z)}\biggr|e^{\beta\omega_0/2}{\cal L}(\omega_0),
\quad
\biggl|\frac{I_0(z)}{2I_1(z)}\biggr|e^{-\beta\omega_0/2}{\cal L}(\omega_0),
\end{eqnarray}
respectively, where ${\cal L}(\omega_0)=\omega_0^2/[1+(\omega_0/\Gamma)^2]=\Gamma^2/[1+(\Gamma/\omega_0)^2]$. We, thus, find that the transmission is significantly reduced when the detuning $\Delta$ equals the first vibrational side-band. Including the remaining contributions to the transmission, i.e. summing over all $n$, provide similar reductions in the transmission at all frequencies $\omega_1=\omega_r+n\omega_0$.

\begin{figure}[b]
\begin{center}
\includegraphics[width=8cm]{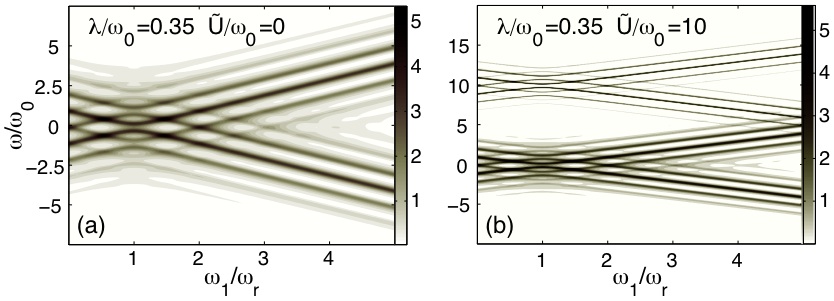}
\end{center}
\caption{(Color online) Local molecular DOS for $\lambda/\omega_0=0.35$, and $\tilde{U}=0$ (a) and $\tilde{U}/\omega_0=10$ (b). Other parameters are as in Fig. \ref{fig-f1}.}
\label{fig-f2}
\end{figure}

\begin{figure}[t]
\begin{center}
\includegraphics[width=8cm]{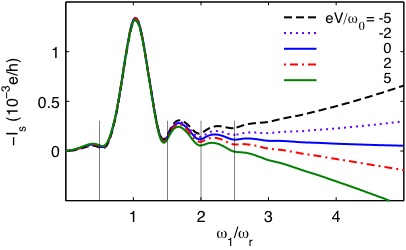}
\end{center}
\caption{(Color online) Bias voltage dependent spin-current for $\lambda/\omega_0=0.35$ and $\tilde{U}=0$. Other parameters as in Fig. \ref{fig-f1}.}
\label{fig-f3}
\end{figure}

We calculate the spin current by solving Eq.~(\ref{EQ:Green_r}) self-consistently.  
The equilibrium spin current through the molecular level is plotted in Fig.~\ref{fig-f1} as function of the rotating frequency $\omega_1$, illustrating the main ESR peak at $\omega_1=\omega_r$ and the vibrational anti-resonances at $\omega_1=\omega_r+n\omega_0$, $n\neq0$ ($\omega_0/\omega_r=1/2$ in the plot). At vanishing correlation energy, the spin-current decreases for increasing coupling strength $\lambda$, which is understood as an effect of the density being distributed among an increasing number of vibrational side-bands for increasing electron-vibron coupling, c.f. Fig. \ref{fig-f2}(a). Increasing spin-current for increasing correlation energy can be explained by the same effect, see Fig. \ref{fig-f2}(b).

It is easy to generalize the above theory to two leads and non-equilibrium conditions. The voltage between the leads is $eV=\mu_L-\mu_R$, and in each lead we have the spin-imbalance such that $\mu_\chi=(\mu_{\chi\up}+\mu_{\chi\down})/2$, $\chi=L,R$. The current $I_{L\sigma}$ for the spin-$\sigma$ current flowing from the left lead into the molecule is written as (with obvious notation)
\begin{eqnarray}
I_{L\sigma}&=&\frac{e}{h}\int
	\Gamma^L\{
	\Gamma^R{\cal T}_c[f_{L\sigma}(\omega)-f_{R\sigma}(\omega)]
\vphantom{\int}
	+\Gamma^L{\cal T}_s
		[f_{L\sigma}(\omega)	
\nonumber\\&&
\vphantom{\int}
		-f_{L\bar\sigma}(\omega)]
	+\Gamma^R{\cal T}_s[f_{L\sigma}(\omega)
	-f_{R\bar\sigma}(\omega)]\}
	d\omega,
\label{eq-sclr}
\end{eqnarray}
where ${\cal T}_c=|G^r_\up\cos^2\phi+G^r_\down\sin^2\phi|^2$, whereas ${\cal T}_s$ is the transmission coefficient given in Eq. (\ref{eq-T}). Here, also $f_{\chi\sigma}(\omega)=f(\omega-\mu_{\chi\sigma})$. The expression for the current in Eq. (\ref{eq-sclr}) is obtained by the observation that the lesser (greater) self-energy in this case is given by
\begin{equation}
\bfSigma^<=i\sum_{\chi\sigma}f_{\chi\sigma}\bfGamma_\sigma^\chi\;,\;\; 
	\bfSigma^>=-i\sum_{\chi\sigma}(1-f_{\chi\sigma})\bfGamma_\sigma^\chi\;.
\end{equation}
We identify the contributions in Eq. (\ref{eq-sclr}) by the first being the usual charge transport as derived by Meir and Wingreen,\cite{meir1992} the second contribution is the one discussed above in Eq. (\ref{eq-sc}), and the third contribution accounts for the spin current between the leads.

The charge current between the leads, $I_c=\sum_\sigma I_{L\sigma}$, becomes
\begin{equation}
I_c=\frac{e}{h}\sum_\sigma\int\Gamma^L\Gamma^R({\cal T}_s+{\cal T}_c)
	[f_{L\sigma}(\omega)-f_{R\sigma}(\omega)]d\omega,
\end{equation}
which is just the sum of the different transmission contributions between the leads. As one would expect, $I_c$ lacks the interference effects that occur in the spin current, which becomes clear by noticing that ${\cal T}_c+{\cal T}_s=|G_\up^r|^2\cos^2\phi+|G_\down^r|^2\sin^2\phi$. The spin-current $I_s=\sum_\sigma\tau^z_{\sigma\sigma}I_{L\sigma}$ is given by
\begin{eqnarray}
I_s&=&\frac{e}{h}\int\Gamma^L\{2\Gamma^L{\cal T}_s[f_{L\up}(\omega)-f_{L\down}(\omega)]
	+\Gamma^R{\cal T}_s[f_{L\up}(\omega)
\nonumber\\&&
\vphantom{\int}
		-f_{R\down}(\omega)
		+f_{R\up}(\omega)-f_{L\down}(\omega)]
	+\Gamma^R{\cal T}_c[f_{L\up}(\omega)
\nonumber\\&&
\vphantom{\int}
		-f_{R\up}(\omega)+f_{R\down}(\omega)-f_{L\down}(\omega)]
	\}d\omega,
\label{eq-ISlr}
\end{eqnarray}
which contains three contributions. The first contribution has the same origin as discussed above in the single medium case; the second contribution accounts for the spin current between the leads;  the third contribution stems from the spin imbalance in the charge current,  which arises from the spin-biased leads.

The non-equilibrium spin current is plotted in Fig.~\ref{fig-f3}, showing its dependence on the bias voltage. For low rotating frequency, the spin current is dominated by transport that is assisted by the rotating magnetic field, c.f. first and second terms in Eq.~(\ref{eq-ISlr}), which provides the main ESR peak and vibrational anti-resonances analogous to the equilibrium case. Increasing frequency $\omega_1$ increases the potential barrier for a molecular level spin-flip. Hence, ac magnetic field assisted transport becomes suppressed, in analogy with the equilibrium situation. The non-equilibrium conditions do, however, enhance tunneling between the leads of electrons that does not undergo spin-flips when in the molecule, i.e. the contribution from the third term in Eq.~(\ref{eq-ISlr}) increases.

We have, for simplicity, neglected effects on the vibrational coherence from the environment, which is justified whenever the dwell time of the localized electrons $\tau_d^{-1}=\Gamma\ll\omega_0$.\cite{armour2001} By studying the vibron mode life-time $\tau_v$, to the second order in the electron-vibron coupling $\lambda$, we find $\tau_v^{-1}\sim\lambda^2\Gamma/[\pi(\omega_0^2+\Gamma^2)]$. In the present study we thus have $\omega_0\tau_v\sim10^2$, for typical electron-coupling strengths and couplings between the local electron with the conduction channel, which justifies the introduced approximations. Assuming a vibron mode $\omega_0\sim1\ \mu$eV, provides a vibron life-time at least in the order of tens of ns, which should be sufficient for measurements.

The herein reported anti-resonances are expected to occur more generally within the ESR set-up. In fact, we have studied the occurrence of the anti-resonances in systems where the local level is coupled to a general system with two or more levels, and we find that the anti-resonances will occur whenever the subsystems have direct interactions with one another. The nature of the interactions may be e.g. tunneling, Coulomb, spin-spin exchange interactions between electrons, or, as discussed in the present paper, fermion-boson interactions.

In summary, we have studied a scheme for electrical detection, using ESR, of vibrational coherence in molecular single electron level trapped near conduction channel. We have shown that the electron-vibron coupling generates anti-resonances in the spin current at frequencies equal to integral numbers of the vibrational mode. The anti-resonances can be explained as interference between opposite spin tunneling electron wave functions traversing different molecular excitations. Observations of the vibrational anti-resonances does not require extremely low temperatures, and since we using realistic parameters for in study, we believe that our findings should be within experimental reach.

J. F. thanks B. Sanyal and P. Souvatzis for useful discussions. This work was supported by the Swedish Research Council (VR)  (J.F.) and by the National Nuclear Security Administration of the U.S. Department of Energy at LANL under Contract No. DE-AC52-06NA25396 and the LANL LDRD Program (J.-X.Z.).

\end{document}